\pdfoutput=1
\documentclass[twocolumn,aps,pra,superscriptaddress,a4paper,amsmath,amssymb,floatfix]{revtex4}
\newcommand{\ket}[1]{\displaystyle{|#1\rangle}}
\newcommand{\bra}[1]{\displaystyle{\langle #1|}}

\usepackage{graphicx}
\usepackage{tabularx}
\usepackage{epstopdf}
\usepackage{color}
\usepackage{amsmath}
\begin{document}
\title{Dissipation induced quantum transport on a finite one-dimensional lattice}

\author{Roland Cristopher F. Caballar}
\email{rfcaballar@up.edu.ph}
\affiliation{National Institute of Physics, College of Science, University of the Philippines, Diliman, 1101 Quezon City}
\author{Bienvenido M.Butanas Jr.}
%\email{bmbutanas@up.edu.ph}
\affiliation{National Institute of Physics, College of Science, University of the Philippines, Diliman, 1101 Quezon City}
\affiliation{Department of Physics, Central Mindanao University, University Town, Musuan, Maramag, Bukidnon, 8710 Philippines}
\author{Vladimir P. Villegas}
%\email{vvillegas@nip.upd.edu.ph}
\affiliation{National Institute of Physics, College of Science, University of the Philippines, Diliman, 1101 Quezon City}
\author{Mary Aileen Ann C. Estrella}
%\email{mary.estrella@outlook.com}
\affiliation{National Institute of Physics, College of Science, University of the Philippines, Diliman, 1101 Quezon City}
\affiliation{Manila Business Consulting Inc., Unit 703 Loyola Heights Condominium, Loyola Heights, Quezon City}

\date{\today}

\begin{abstract}
We construct a dissipation induced quantum transport scheme by coupling a finite lattice of $N$ two-level systems to an environment with a discrete number of energy levels. With the environment acting as a reservoir of energy excitations, we show that the coupling between the system and the environment gives rise to a mechanism for excited states of the system to be efficiently transported from one end of the lattice to another. We also show that we can adjust the efficiency of the quantum transport scheme by varying the spacing between energy levels of the system, by decreasing the ground state energy level of the environment, and by weakening the coupling between the system and the environment. A possible realization of this quantum transport scheme using ultracold atoms in a lattice coupled to a reservoir of energy excitations is briefly discussed at the end of this paper.  
\end{abstract}

\maketitle

\section{Introduction}

Dissipation in quantum mechanics has, in recent times, attracted an increasing amount of interest. This is primarily due to its role in inducing decoherence in a quantum system which is coupled to an environment \cite{breuer, weiss}. In this context, dissipation can be viewed as a bane in quantum mechanics, and efforts have been made to reduce dissipative effects due to correlation between a quantum system and the surrounding environment \cite{breuer, nielsen}.

However, recent work has shown that dissipation can be used as a resource in quantum mechanics, wherein it drives the evolution of a quantum system towards a unique steady state \cite{breuer}. In particular, if a system is coupled to an environment in the manner of an open quantum system, then with the proper choice of environment, and by adjusting the strength of coupling between the system and the environment, the resulting time evolution equation for the system will have a unique steady state. The existence of this unique steady state can then be attributed to dissipative effects due to the coupling between the system and the environment. As such, one can then use dissipation as a resource in quantum computation and quantum state preparation, as shown in Refs. \cite{verstraete, diehl, diehl2, dallatorre, caballar}.

Aside from quantum computation and quantum state preparation, dissipation can also be used as a resource in quantum state transport. An illustration of how this can be done was carried out by Rebentrost et. al. \cite{rebentrost}, wherein they considered an interacting $N$-body system in the presence of a single excitation. They were able to show that by coupling this system with a fluctuating environment, the quantum transport of excited energy states can be enhanced, with the efficiency of the process dependent on the energy mismatch between states and the hopping terms in the system Hamiltonian. This quantum transport scheme has been applied to the analysis of electronic energy transfer in photosynthetic structures \cite{palmeiri,ishizaki,fassioli,panitchayangkoon,cai,ishizaki2}, in non-Markovian open quantum systems \cite{liang, sweke}, as well as in the analysis of quantum transport in various systems \cite{mulken, caruso, sinayskiy, antezza}.

Another possible mechanism for efficient dissipation-assisted quantum transport was provided in Refs. \cite{attal,attal2, sinayskiy2}, which makes use of open quantum random walks. In this mechanism, a system with internal and spatial degrees of freedom is coupled to an environment, with the coupling between the system and the environment causing the system to undergo a quantum random walk. Open quantum random walks have been shown to obey a central limit theorem \cite{konno,attal3}, which implies that quantum systems undergoing open quantum random walks will evolve towards a unique steady state.

Having shown that dissipation can be treated as a resource in quantum mechanics, and that it can be used to enhance and create efficient quantum transport mechanisms, we then ask if it is possible to create other dissipation induced quantum transport mechanisms. In this paper, we show that it is possible by considering a system comprised of a lattice of two-level systems coupled to an environment with a discrete number of energy levels. We show that if the system and the environment are weakly coupled to each other, it is possible to create an efficient dissipation induced quantum transport scheme for excited states of the system from one end of the lattice to the other. Maximum efficiency can be achieved if the number of energy levels present in the environment is roughly of the same order of magnitude as the system's, if the spacing between energy levels of the system is relatively large and if the ground state energy of the environment is much less than that of the system's. 

The rest of the paper is divided into the following sections. Section 2 gives a general description of the system and environment in terms of their respective Hamiltonians, and specifies as well the form of the Hamiltonian describing their interactions. Section 3 outlines the derivation of the master equation describing the dynamics of the system, while Section 4 provides a description of the dynamics of the system by examining the properties of the numerical solution of the master equation of the system. We summarize our results in Section 6.

\section{A lattice of two-level systems coupled to an environment}

The system $S$ considered in this paper consists of a one-dimensional lattice of two-level systems. The Hilbert space corresponding to this system is given as $\mathcal{H}_{S}=\mathcal{H}_{S,int}\otimes\mathcal{H}_{S,pos}$, where $\mathcal{H}_{S,int}$ and $\mathcal{H}_{S,pos}$ are the subspaces of the system's Hilbert space corresponding to the system's energy and position, respectively. In this Hilbert space, the system's Hamiltonian has the following form:

\begin{equation}
\mathrm{H}_{S}=\sum_{n=1}^{2}\sum_{j=1}^{N}\varepsilon_{n}\hat{a}^{\dagger}_n \hat{a}_n\otimes\ket{j}\bra{j},
\label{hamsys}
\end{equation}
where $\ket{j}$ is a basis vector in $\mathcal{H}_{S,pos}$ corresponding to node $j$ in the lattice, which is finite, 1-dimensional and has a total of $N$ nodes. Also, the operator $\hat{a}_n$ is the annihilation operator defined in the Hilbert space $\mathcal{H}_{S,int}$ corresponding to the energy level $\varepsilon_n$ for the system's internal degrees of freedom. 

Let the system $S$ be coupled to an environment $B$, which has a Hamiltonian $\mathrm{H}_B$ whose explicit form is defined in a Hilbert space $\mathcal{H}_B$ as 

\begin{equation}
\mathrm{H}_{B}=\sum_{k=1}^{M}E_{k}\hat{b}^{\dagger}_{k}\hat{b}_k,
\label{hamenv}
\end{equation}
where $\hat{b}_k$ is the annihilation operator, defined in $\mathcal{H}_B$, corresponding to the energy level $E_k$ of the environment $B$. 

To describe the interaction between the system and the environment, we assume that the coupling between the system and the environment is linear in system operators defined in $\mathcal{H}_S$ and $\mathcal{H}_B$. Furthermore, we let those system operators be $\hat{a}_{n}\otimes\ket{j}\bra{j}$ and $\hat{b}^{\dagger}_{k}$ which are defined in $\mathcal{H}_S$ and $\mathcal{H}_B$ respectively. Then the interaction between the system and environment is described by the following Hamiltonian \cite{breuer}:

\begin{align}
\mathrm{H}_{SB}=&\sum_{k}\sum_{n=1}^{2}\sum_{j=1}^{N}g_{nkj}\hat{a}_{n}\otimes\ket{j+1}\bra{j}\otimes\hat{b}^{\dagger}_{k}\nonumber\\
&+g^{*}_{nkj}\hat{a}^{\dagger}_{n}\otimes\ket{j}\bra{j+1}\otimes\hat{b}_{k},
\label{hamtot}
\end{align}
where $g_{nkj}$ is the coupling constant describing the strength of coupling between the system $S$ and the environment $B$. We evolve the coupling Hamiltonian over time, making use of the evolution equation

\begin{displaymath}
\mathrm{H}_{SB}(t)=\mathrm{exp}\left(-\frac{i}{\hbar}(\mathrm{H}_{S}+\mathrm{H}_{B})t\right)\mathrm{H}_{SB}\mathrm{exp}\left(\frac{i}{\hbar}(\mathrm{H}_{S}+\mathrm{H}_{B})t\right).
\end{displaymath}
In doing so, we obtain the following expression:

\begin{align}
\mathrm{H}_{SB}(t)=\sum_{n,k}\sum_{j=1}^{N}&\mathrm{e}^{-\frac{i}{\hbar}\left(\varepsilon_{n}-E_{k}\right)t}g_{nkj}\hat{a}_{n}\otimes\ket{j+1}\bra{j}\otimes\hat{b}^{\dagger}_{k}\nonumber\\
&+\mathrm{e}^{\frac{i}{\hbar}\left(\varepsilon_{n}-E_{k}\right)t}g^{*}_{nkj}\hat{a}^{\dagger}_{n}\otimes\ket{j}\bra{j+1}\otimes\hat{b}_{k}.
\label{timeevohamtot}
\end{align}

From the form of $\mathrm{H}_{SB}$, we can then see that the coupling between the system and the environment induces a form of quantum transport of excitations in the system from one lattice site to another, and in doing so either raises or lowers the energy of the environment. This quantum transport process of excitations can be described more explicitly using a quantum master equation for the system, which will be derived in the next section. 

\section{Master equation for the lattice of two-level systems coupled to an environment}
Allowing the system to interact with the environment means that it is now an open quantum system, whose dynamics are described by a master equation which specifies the time evolution of the density matrix describing the system. This master equation can be obtained by making use of the integral form of the von Neumann master equation in the interaction picture, and by making use of the Born approximation as well as the assumption that the initial state is a product state \cite{breuer}. In doing so, we will then obtain the Redfield equation, whose general form is given by
\begin{equation}
\frac{d}{dt}\rho_{S}(t)=-\int_{0}^{t}ds\left[\mathbf{H}_{SB}(t),\left[\mathbf{H}_{SB}(s),\rho_{S}(t)\otimes\rho_{B}\right]\right].
\label{redfieldgen}
\end{equation}

Here, $\rho_{S}(t)$ is the density matrix describing the system $S$. We make use of the Redfield equation rather than the Born-Markov equation to describe the dynamics of the open quantum system because we are interested in determining the dynamics of the system over intermediate timescales, rather than over long timescales. Such timescales are more realistic and experimentally realizable, which implies that the resulting master equation will be of greater use in experimental investigations of the open quantum system. In deriving this equation, we make use of the Born approximation, which states that the total density matrix of the system coupled to the environment has the form

\begin{equation}
\rho(t)=\sum_{j=1}^{N}\rho_{S}(t)\otimes\ket{j}\bra{j}\otimes\rho_{B},
\label{densmattime}
\end{equation}
where $\rho_{S}(t)$ and $\rho_B$ are the density matrices, defined in the Hilbert spaces $\mathcal{H}_{S,int}$ and $\mathcal{H}_B$, respectively, describing the state of the internal degrees of freedom of the system and of the environment at node $j$ of the lattice and at the instant of time $t$.

Now for a system undergoing a quantum walk, the density matrix $\rho_{S}(t)$ describing the system at time $t$ can be written as 

\begin{equation}
\rho_{S}(t)=\sum_{n,j}\rho_{n,j}(t),
\end{equation}

where $\rho_{n,j}(t)$ describes the state of the system at energy level $n$ and node $j$ in the lattice. Inserting equations \ref{timeevohamtot} and \ref{densmattime} into equation \ref{redfieldgen}, we then obtain the following expression:

\begin{align}
&\sum_{j,n}\frac{d}{dt}\rho_{n,j}(t)\otimes\ket{j}\bra{j}=\nonumber\\
&\sum_{j=1}^{N}\sum_{n}\left(-i\Gamma_{nj}(t)\left[\hat{a}_{n}\hat{a}^{\dagger}_{n},\rho_{S,j}(t)\right]\otimes\ket{j}\bra{j}\right.\nonumber\\
&\left.+\gamma_{nj}(t)\left(2\hat{a}_{n}\rho_{n,j+1}(t)\hat{a}^{\dagger}_{n}-\left\{\hat{a}^{\dagger}_{n}\hat{a}_{n},\rho_{n,j}(t)\right\}\right)\right)\otimes\ket{j}\bra{j}.
\label{masteqdyn}
\end{align}
Details about the derivation of this master equation are given in the appendix of this paper. Here, the coefficients $\Gamma_{nj}(t)$ and $\gamma_{nj}(t)$ have the following form:

\begin{align}
&\Gamma_{nj}(t)=\sum_{k}\frac{\hbar}{\epsilon_{n}-E_{k}}|g_{kn}|^{2}\left(1-\cos\left(\frac{\epsilon_{n}-E_{k}}{\hbar}t\right)\right),\nonumber\\
&\gamma_{nj}(t)=\sum_{k}\frac{\hbar}{\epsilon_{n}-E_{k}}|g_{kn}|^{2}\sin\left(\frac{\epsilon_{n}-E_{k}}{\hbar}t\right).
\label{masteqcoeffs}
\end{align}

The resulting master equation is block diagonal in the system Hilbert space $\mathcal{H}_{S}=\mathcal{H}_{S,int}\otimes\mathcal{H}_{S,pos}$ just like the density matrix of the system. However, the Lindblad operator of the master equation, given by
\begin{align}
&\mathcal{L}_{n,k,j}(\rho_{S,j}(t))\nonumber\\
&=-\gamma_{nkj}\left(2\hat{a}^{\dagger}_{n}\rho_{S,j+1}(t)\hat{a}_{n}-\left\{\hat{a}_{n}\hat{a}^{\dagger}_{n},\rho_{S,j}(t)\right\}\right),
\end{align}
involves two internal states of the system, namely the internal state of the system at lattice site $j+1$ and the internal state of the system at lattice site $j$. This, together with the time dependence of the coefficients $\gamma_n$ of the Lindblad operator, signifies that $\mathcal{L}_{n,j}$ is not in Lindblad form, which implies that the time evolution of the internal states of the system is non-Markovian. We will explore the consequences of this non-Markovian behavior of the system in the next section as we examine the dynamics of the system by numerically solving the master equation. 

\section{Dynamics of the system}

Having derived the master equation describing the dynamics of the system, we now solve it numerically to enable us to analyze the dynamical behavior of the system. In doing so, we will also be able to examine the quantum transport mechanism described by the system-environment interaction Hamiltonian given in section 2, and determine the efficiency of such a process. In our analysis, we make use of natural units.

For the initial state, we assume that it is localized at node $j=1$, and is then given as $\rho(0)=\rho_{S}(0)\otimes\ket{1}\bra{1}\otimes\rho_{B}$, where $\rho_{S}(0)$ is the density matrix describing the initial internal state of the system. Furthermore, we assume that the system and the environment are weakly coupled to each other, i. e. $\gamma_{nk}<<1$ and $\Gamma_{nk}<<1$. In all computations, we assume that the initial internal state $\rho_{S}(0)$ of the system localized at node $j=1$ is the ground energy state of the system. To analyze the dynamics of the system, we then compute for the probability that a node $\ket{j}$ is occupied at time $t$ as follows:

\begin{equation}
P_{j}(t)=\left|\mathrm{Tr}\left(\bra{j}\rho_{S}(t)\ket{j}\right)\right|,
\label{occprob}
\end{equation} 
where the trace is taken over the internal state $\rho(t)$ occupying node $j$ at time $t$.

\begin{figure}[htb]
	\includegraphics[width=0.5\textwidth,height=0.3\textheight]{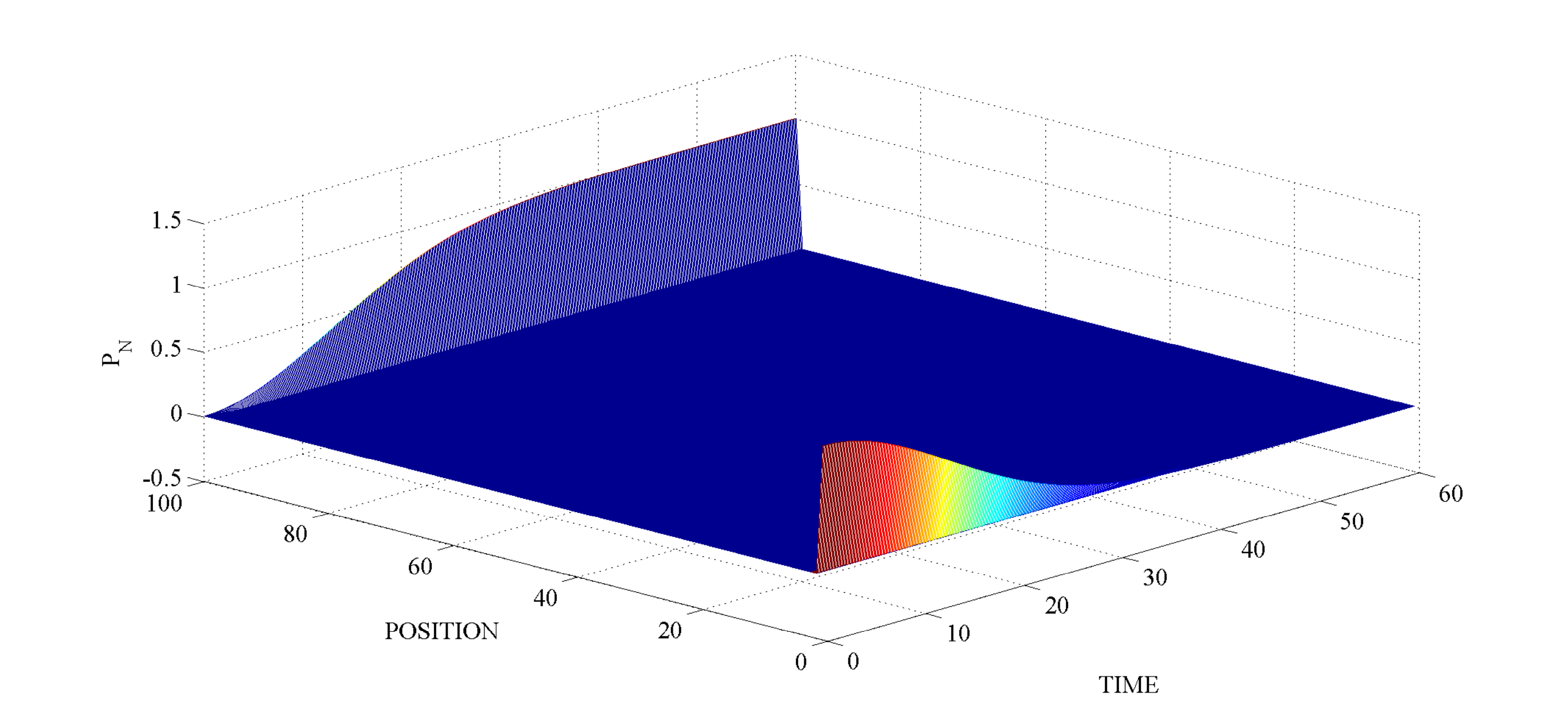}
	
	\caption{Plot of the occupation probability $P_{j}(t)$ of each node in the lattice from $j=1$ to $j=100$ from $t=0$ to $t=60$, with time step $\Delta t = 0.01$, for a 2-level system with a large gap between its ground and excited energy levels coupled to an environment with 5 energy levels, a ground energy level lower than the system's, and large gaps between each of its energy levels.}
	\label{OccProbAllNodesAllTimes}
\end{figure}

By computing the occupation probability $P_{j}(t)$ for various nodes, we find that an optimal quantum transport scheme will result from the coupling of the two-level system to the environment if the spacing between energy levels in the system and the environment are large $(\Delta_E = E_{e}-E_{g}>>E_{g})$, if the number of energy levels in the environment is small, and if the ground state energy of the environment is much smaller than the ground state energy of the system $(E_{g,B}<<E_{g,S})$. This is shown in Fig. \ref{OccProbAllNodesAllTimes}, which is a plot of the occupation probability $P_{j}(t)$ taken over all nodes at each time $t$. The plot shows that initially, the state is localized at the origin, but as it evolves, the probability that it will be found at the end node increases while the probability that it is found at any other node decreases. Eventually, the state is localized at the end node of the lattice at an instant of time $t<N$, where $N$ is the number of nodes in the lattice. Hence, if we define the speed of transport of the state as

\begin{equation}
v_{N}=t_{max,N}/N,
\label{speedtrans}
\end{equation}
where $t_{max,N}$ is the instant of time when the occupation probability at the end node of the lattice reaches its maximum value, then for quantum transport of a state from one end of a lattice to another to be efficient, $v_{N}<1$, which is exactly what is observed for the quantum transport scheme due to the coupling between the system and the environment considered in this paper, for the conditions specified above. 

\begin{figure}[htb]
	\includegraphics[width=0.53\textwidth,height=0.3\textheight]{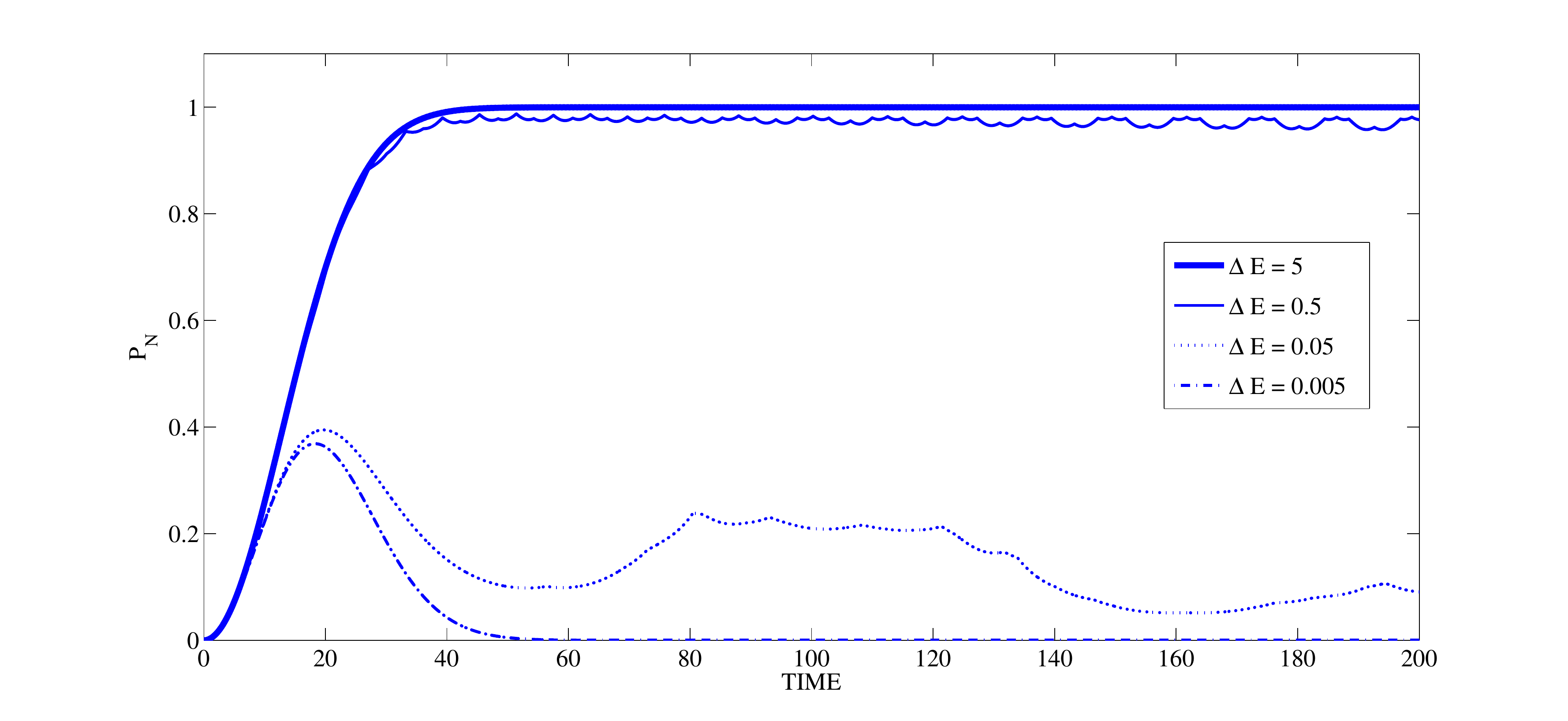}
	
	\caption{Plot of the probability that the end node of a 100-node 1-D lattice is occupied as a function of time, for varying energy gaps between the ground and excited states of the system as indicated in the plot.}
	\label{OccProbVaryingSystemEnergyGap}
\end{figure}

The effect of decreasing the spacing between the ground and excited energy levels of the system is shown in Fig. \ref{OccProbVaryingSystemEnergyGap}. In particular, we see that for a relatively large energy gap between the ground state and excited state of the system, the coupling between the system and the environment will drive the system towards a steady state which occupies the end node of the system beginning at an instant of time less than the number of nodes in the lattice. This signifies that the transport process due to the coupling between the system and the environment is efficient, since the amount of time it takes for the end node of the lattice to be occupied is less than the total number of nodes in the lattice. On the other hand, the smaller the energy gap between the ground state and excited state of the system, the smaller is the maximum value of the probability that the end node of the lattice is occupied. 

Furthermore, as Fig. \ref{OccProbVaryingSystemEnergyGap} shows, as the energy gap between the ground and excited states of the system decreases, the more likely that the steady state of the system will not be one that occupies the end node of the lattice. This implies that decreasing the energy gap between the system's energy levels also decreases the efficiency of the quantum transport process due to the coupling between the system and the environment.

\begin{figure}[htb]
	\includegraphics[width=0.53\textwidth,height=0.3\textheight]{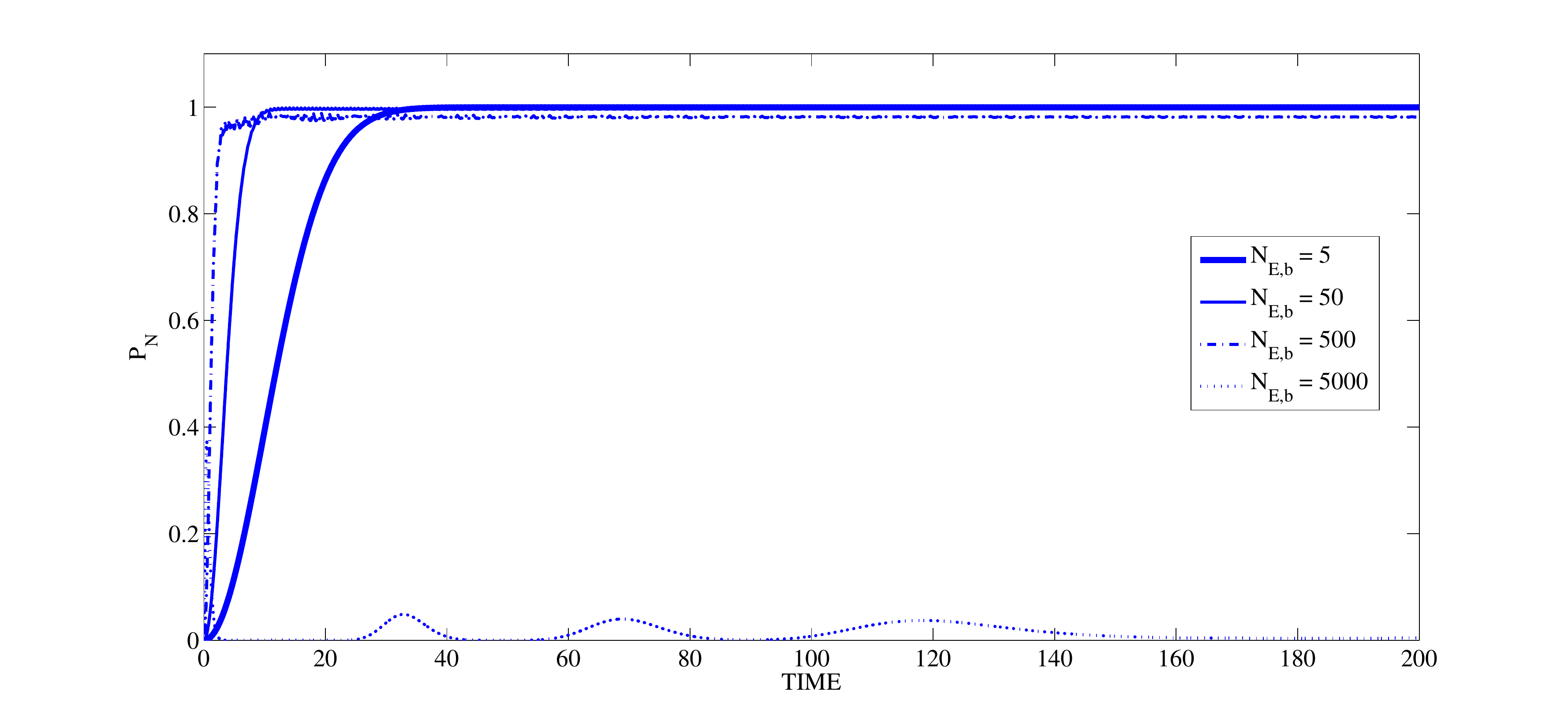}
	
	\caption{Plot of the probability that the end node of a 100-node 1-D lattice is occupied as a function of time, for varying number of energy levels in the environment, as indicated in the plot.}
	\label{OccProbVaryingEnvtEnergyLevelNum}
\end{figure}

As for the effect of increasing the number of energy levels available in the environment, Fig. \ref{OccProbVaryingEnvtEnergyLevelNum} shows that increasing the number of energy levels decreases the efficiency of the quantum transport process due to the coupling between the system and the environment. In particular, as the number of energy levels in the environment increases, the instant when the probability that the endpoint of the lattice is occupied is at its maximum occurs at an earlier time. However, the maximum value of this probability will also decrease. This then implies a decrease in the efficiency of the quantum transport process, since there is a nonzero probability that other nodes in the lattice other than the endpoint are occupied.

While the number of energy levels present in the environment has an effect on the efficiency of the quantum transport scheme, the spacing between energy levels in the environment apparently has no effect whatsoever. Therefore, an environment with a small number of discrete energy levels of arbitrary spacing from each other, coupled to a lattice of two-level systems, will create an efficient quantum transport scheme from one end of the lattice to the other.

\begin{figure}[htb]
	\includegraphics[width=0.53\textwidth,height=0.3\textheight]{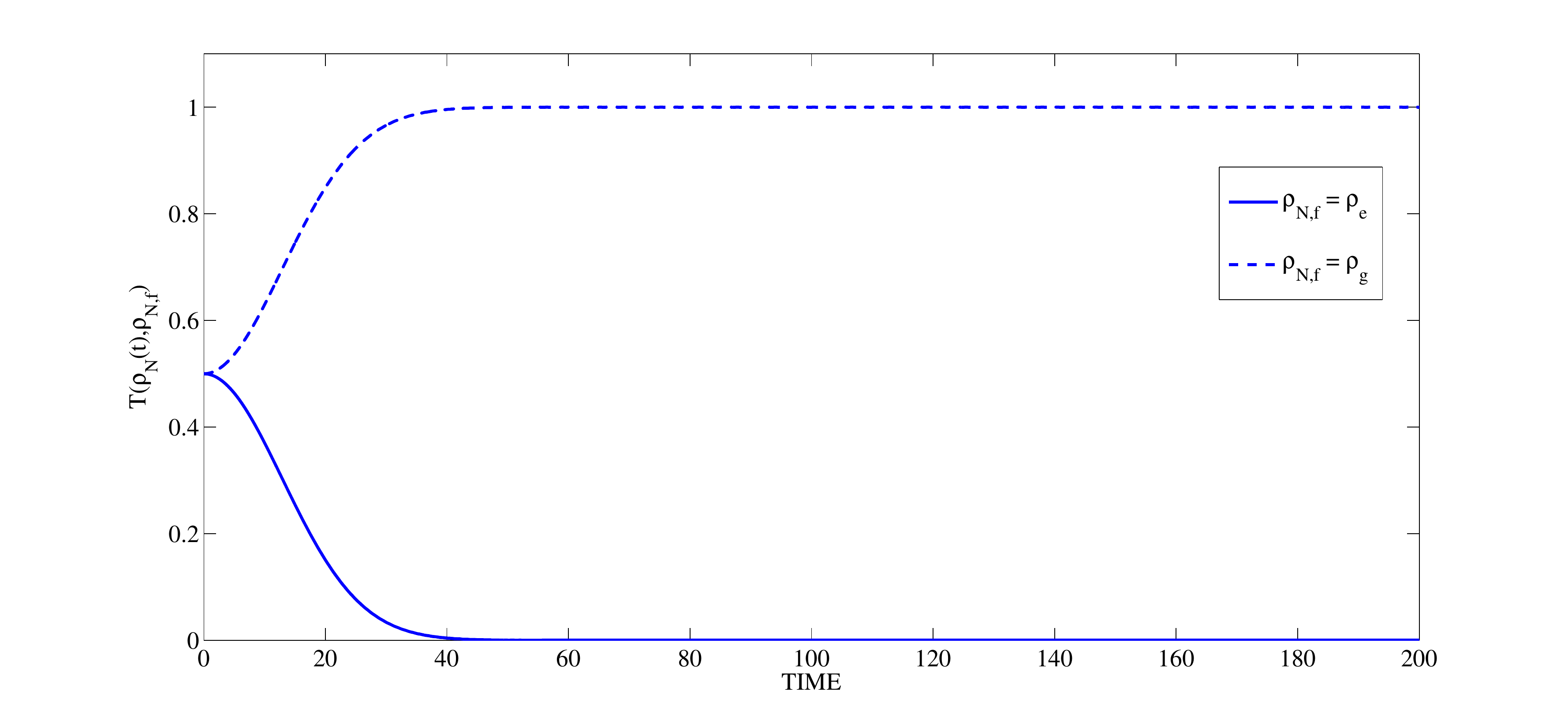}
	
	\caption{Plot of the trace distance between the time-evolved state given by the density matrix $\rho_{N}(t)$ at the end of a 100-node 1-D lattice and a target final state $\rho_{N,f}$. The target final states at the end of the lattice are either the ground state $\rho_{g}$ or excited state $\rho_e$ of the system.}
	\label{TraceDistGrndandExci}
\end{figure}

\begin{figure}[htb]
	\includegraphics[width=0.53\textwidth,height=0.3\textheight]{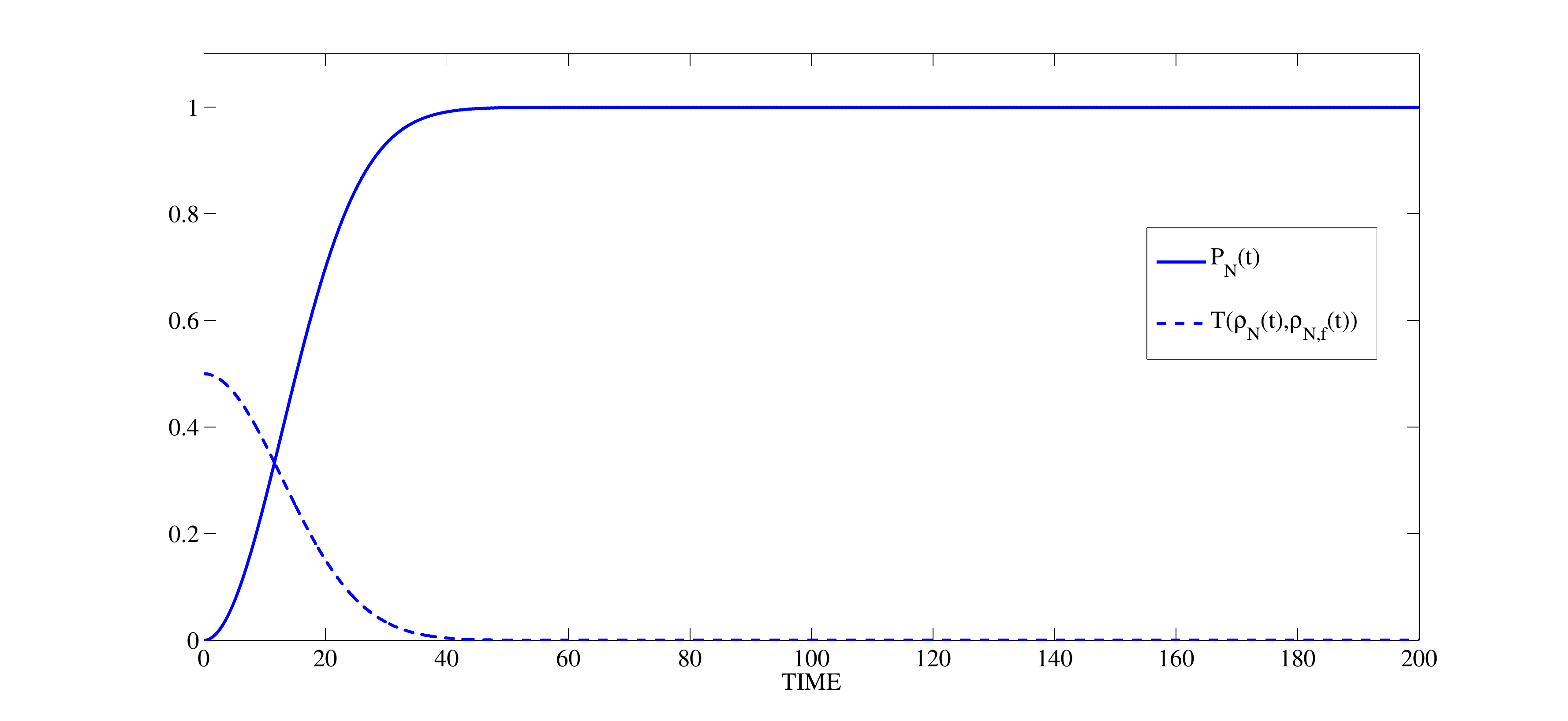}
	
	\caption{Plot of the occupation probability (solid line) and the trace distance between the time-evolved state given by the density matrix $\rho_{N}(t)$ (dashed line) at the end of a 100-node 1-D lattice and a target final state $\rho_{N,f}$. The target final state at the end of the lattice is the excited state $\rho_e$ of the system.}
	\label{OccProbandTraceDist}
\end{figure}

Finally, there is the question of what exactly is the state that is transported to the end node of the lattice. To determine what state is transported to the end of the lattice, we compute for the trace distance between the state at the end of the lattice, as given by the density matrix $\rho_{N}(t)=\rho(t)\otimes\ket{N}\bra{N}$, and a desired final state given by $\rho_{N,f}=\rho_{f}\otimes\ket{N}\bra{N}$. The trace distance is defined as

\begin{equation}
\mathrm{T}\left(\rho_{N}(t),\rho_{N,f}\right)=\frac{1}{2}\mathrm{Tr}\left(\sqrt{\left(\rho_{N}(t)-\rho_{N,f}\right)^2}\right)=\frac{1}{2}\sum_{j}\left|\lambda_{j}\right|,
\label{tracedist}
\end{equation}
where $\lambda_j$ are the absolute values of the eigenvalues of the Hermitian matrix $\rho_{N}(t)-\rho_{N,f}$. As shown in Fig. \ref{TraceDistGrndandExci}, the trace distance drops off to zero if $\rho_{N,f}$ is an excited energy state of the system, while it rises to one if $\rho_{N,f}$ is a ground energy state of the system. This signifies that if the initial state of the system is the ground energy state, then the state transported to the end of the lattice is the excited energy state of the system. This is further illustrated in Fig. \ref{OccProbandTraceDist}, wherein the occupation probability at the end of the lattice reaches its maximum value at the same instant that the trace distance between $\rho_{N}(t)$ and $\rho_{N,f}$ equals zero, signifying that the state that is transported at the end of the lattice is indeed an excited energy state of the system.

\begin{figure}[htb]
	\includegraphics[width=0.53\textwidth,height=0.3\textheight]{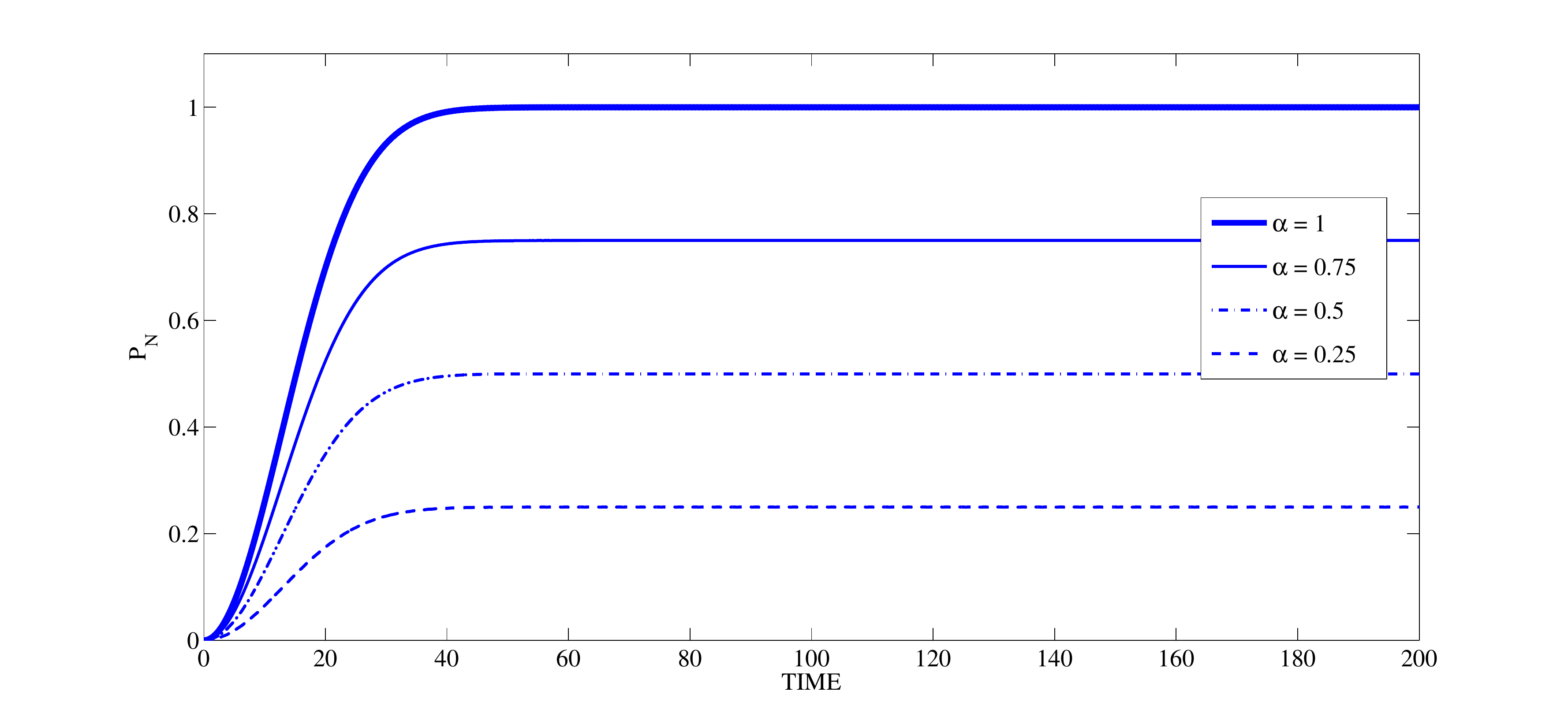}
	
	\caption{Plot of the occupation probability at the end of a 100-node 1-D lattice, for varying initial states of the form $\rho(0)=\alpha\rho_{g}+\beta\rho_{e}.$}
	\label{OccProbVaryingInitState}
\end{figure}

We note that the calculations and results obtained up to this point in this section depend on the initial state of the system being in the ground energy state. If, however, the initial state is not the ground state, then the efficiency of the quantum transport scheme will be greatly affected. In particular, as shown in Fig. \ref{OccProbVaryingInitState}, the probability that the endpoint of the lattice will be occupied decreases if the initial state localized at the beginning of the lattice has the form
\begin{equation}
\rho_{S}(0)=\alpha\rho_{g}+\beta\rho_{e},
\label{initstategenform}
\end{equation} 
where $\rho_g$ and $\rho_e$ are the density matrices corresponding to the ground and excited states of the system. In fact, the maximum value of $P_{N}$ will be equal to the coefficient $\alpha$ in the expression for the initial state $\rho_{S}(0)$ as given by Eq. \ref{initstategenform}.

\section{Comparison of Dynamics of the System with that of a similar Markovian system}

Having analyzed the dynamics of the open quantum system, we now compare it to the dynamics of a similar system whose master equation was obtained using the Born-Markov equation. In this approximation, the timescale over which the system varies is much smaller than the timescale over which the environment varies. The system, environment and interaction Hamiltonians of this open quantum system are given by Eqs. \ref{hamsys}, \ref{hamenv} and \ref{hamtot} respectively, and the derivation of its master equation will follow lines similar to those given in the Appendix. However, the point of departure in deriving the open quantum system's master equation comes after Eq. \ref{doublecommtrenvo}. This is because the double commutator is integrated over a semi-infinite time interval $[0,\infty)$ instead of over a finite time interval $[0,t]$. In doing so, we obtain the following form of the master equation for this system:

\begin{align}
&\sum_{j,n}\frac{d}{dt}\rho_{n,j}(t)\otimes\ket{j}\bra{j}=\nonumber\\
&\sum_{j=1}^{N}\sum_{n}\left(-i\Gamma_{nj}\left[\hat{a}_{n}\hat{a}^{\dagger}_{n},\rho_{S,j}(t)\right]\otimes\ket{j}\bra{j}\right.\nonumber\\
&\left.+\gamma_{nj}\left(2\hat{a}_{n}\rho_{n,j+1}(t)\hat{a}^{\dagger}_{n}-\left\{\hat{a}^{\dagger}_{n}\hat{a}_{n},\rho_{n,j}(t)\right\}\right)\right)\otimes\ket{j}\bra{j}.
\label{masteqdynmarkov}
\end{align}

\begin{figure}[htb]
	\includegraphics[width=0.53\textwidth,height=0.3\textheight]{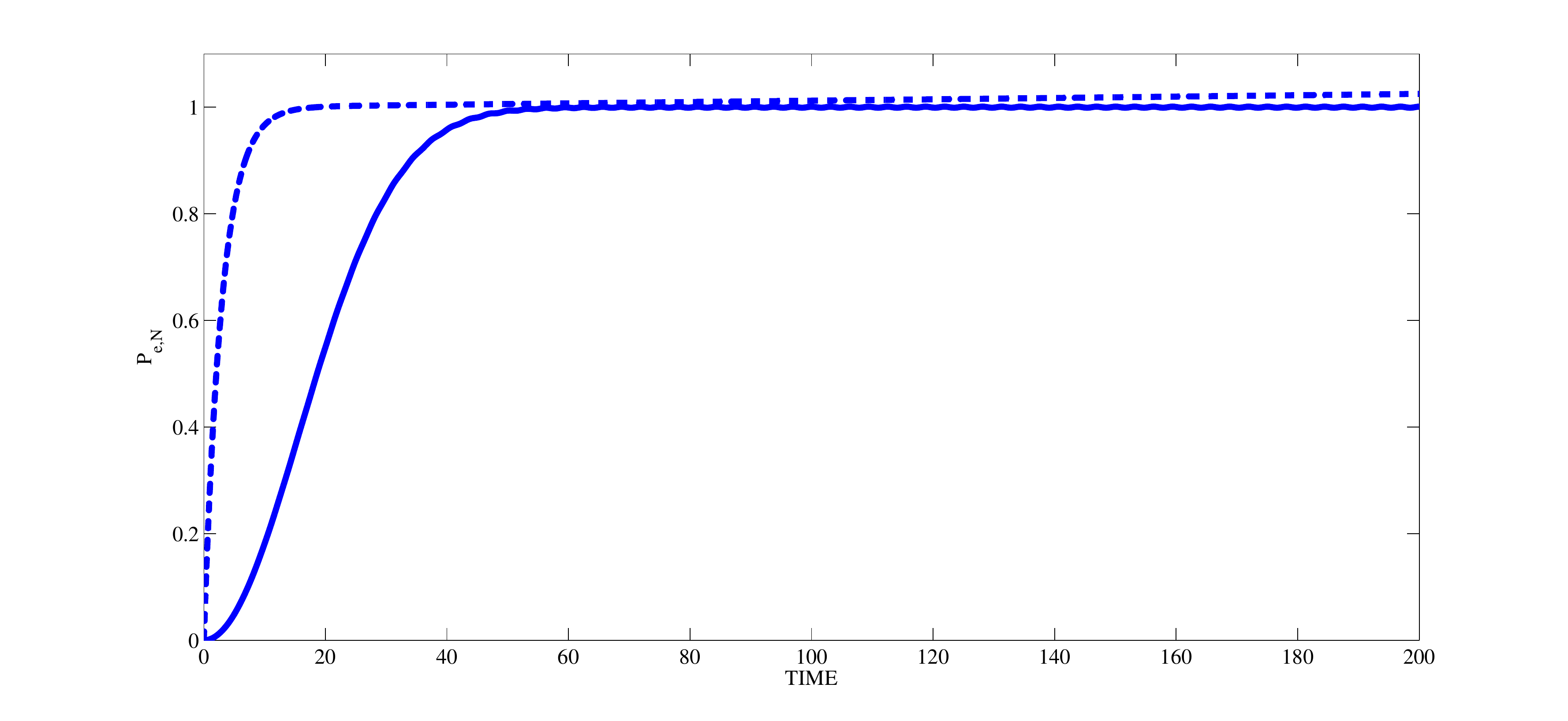}
	
	\caption{Plot of the probability that the end node of a 100-node 1-D lattice is occupied as a function of time, for an open quantum system described by Eq. \ref{masteqdyn} (solid line) and \ref{masteqdynmarkov} (dashed line) respectively.}
	\label{OccProbNodeRedfieldvsMarkov}
\end{figure}

Here, the coefficients $\Gamma_{nj}$ and $\gamma_{nj}$ are time-independent, and have the explicit form

\begin{align}
&\Gamma_{nj}=\sum_{k}|g_{kn}|^{2}\int_{0}^{\infty}ds\;\sin\left(\frac{\epsilon_{n}-E_{k}}{\hbar}s\right),\nonumber\\
&\gamma_{nj}=\sum_{k}|g_{kn}|^{2}\int_{0}^{\infty}ds\;\cos\left(\frac{\epsilon_{n}-E_{k}}{\hbar}t\right).
\label{masteqcoeffsmarkov}
\end{align}
We note that unlike in the master equation given by Eq. \ref{masteqdyn}, the coefficients of the master equation given by Eq. \ref{masteqdynmarkov} are time-independent. As we have done in the previous section, to analyze the dynamics of the system described by Eq. \ref{masteqdynmarkov}, we compute for the probability that, if the system's initial state is an excited energy state localized at node $\ket{1}$, which is the starting point of the one-dimensional lattice in which the system moves, the node $\ket{j}$ is occupied in this system at time $t$, with this probability given by Eq. \ref{occprob}. We then compare this to the probability that the same node is occupied in an open quantum system which is described by Eq. \ref{masteqdyn} and whose initial state is also an excited energy state localized at node $\ket{1}$. Our results are summarized in Fig. \ref{OccProbNodeRedfieldvsMarkov}. The figure shows that for both the system described by Eq. \ref{masteqdyn} and the system described by Eq. \ref{masteqdynmarkov}, the probability that the node $\ket{N}$, which is the endpoint of the one-dimensional lattice in which these systems are evolving, will attain maximal values at instants of time $t<N$, with these maximal values  implying that both systems can be used for efficient quantum transport of excited states from one end of a one-dimensional lattice to another. However, we also find that the instant when the open quantum system described by Eq. \ref{masteqdynmarkov} attains a maximal value for Eq. \ref{occprob} at $\ket{N}$ occurs much earlier than the instant when the open quantum system described by Eq. \ref{masteqdyn} attains a maximal value for Eq. \ref{occprob} at $\ket{N}$. This suggests that if one makes use of the Born-Markov approximation to obtain the master equation describing the dynamics of the open quantum system considered in this paper, then that system can be used for dissipation-assisted efficient quantum transport of an excited state from one end of a one-dimensional lattice to another, and that such a quantum state transport scheme will be much more efficient than one which makes use of the same system but which does not make use of the Born-Markov approximation. However, we note that the Redfield equation is much more general than the Born-Markov approximation, since the former does not make any requirements regarding the timescales over which the system and the environment vary. As such, the use of the Born-Markov approximation will be able to provide a qualitiative description of the efficiency of the dissipation induced quantum transport scheme considered in this paper, but the use of the Redfield equation will allow us to consider more systems that can be used to physically realize this quantum transport mechanism.

\section{Analysis and Discussion}
From the previous section, we found that it is possible to construct a dissipation induced quantum transport mechanism for excited states of a two-level system in a lattice by weakly coupling it to an environment with a small number of discrete energy levels. The resulting quantum transport mechanism will be efficient, in that the amount of time it will take to transport the excited energy state of the system from one end of the lattice to the other will be less than the number of nodes in the lattice. 

Such a quantum transport mechanism has the advantage of eliminating active control over the system throughout the process. Rather, it is the interaction between the system and the environment that allows the quantum transport scheme to be carried out. All that is necessary to allow the quantum transport scheme to be carried out is to prepare and localize the initial state at one end of the lattice.

We note that the quantum transport scheme for the system described in this paper is optimized for transporting excited states from one end of the lattice to another, with the initial state being the ground state for the system. However, the state that is transported to the other end is orthogonal to the initial state of the system localized in one end. There is no other mechanism present in the system other than its coupling with the environment to explain why the initial state localized in one end of the lattice is different from the state transported to the other end. As such, not only does the coupling between the system and the environment induce transportation of excited states from one end of the lattice to another, it also excites the initial state of the system from the ground to the excited energy level. Hence, a full description of the dissipation induced quantum transport scheme due to the coupling between the system and the environment described in this paper can be given as follows: if the initial state of the system is in the ground state and is localized at one end of the lattice, the interaction between the system and the environment will first raise the energy of the initial state to the excited level, then will cause the excited state to be transported to the other end of the lattice at an instant of time less than the number of nodes in the lattice. 

The results obtained in the previous section also imply that even if the initial state of the system is not entirely in the ground state, but rather is a superposition of the system's ground and excited states, coupling the system to the environment described in this paper will still cause an excited state of the system to be transported from one end of the lattice to the other. However, the efficiency of this quantum transport scheme in transporting the system's excited state from one end of the lattice to the other will be reduced, since the probability that the state transported to the end of the lattice is an excited state will be less than one. Instead, what is transported to the other end of the lattice is the ground state of the system. Nevertheless, the coupling between the system and the environment still results in a dissipative quantum transport scheme for either ground or excited energy states of the system, which first changes the energy of the initial state before transporting the resulting excited or deexcited energy state of the system from one end of the lattice to the other end.

Finally, we note that the qualitative behavior of this dissipative quantum transport mechanism can be obtained using a master equation obtained using the Born-Markov approximation. Such an approximation will give us a much simpler form of the master equation, since its coefficients will be time-independent. However, it imposes more constraints onto the system, in particular requiring that the environment varies much more slowly over time than the system, a requirement which may limit the types of systems that can realize this quantum transport mechanism. Nevertheless, the use of the Born-Markov approximation to describe this quantum transport scheme is still useful, due to the simplicity of the resulting master equation that allows for quicker and easier analysis of the dynamical behavior of the mechanism, as well as the qualitative similarity of the dynamical behavior described by this equation with that described by the Redfield equation for this same quantum transport scheme.

However, there are still issues that need to be resolved with this quantum transport scheme, foremost of which is its physical realizability. In particular, there is the question of what the explicit form of the system and the environment that can be used to realize this quantum transport scheme. One possible physical realization of this quantum transport scheme can be accomplished by taking an ensemble of two-level ultracold atoms confined to a lattice as our system, and weakly coupling these two-level atoms in the lattice to a Bose-Einstein Condensate (BEC), whose ground state energy is much less than the ground state of the two-level atoms in the lattice. We note that BECs have been proposed as environments coupled to particular systems before, in particular in Refs. \cite{diehl, verstraete, diehl2, caballar}. They are able to absorb and emit excitations from and into the system to which they are coupled, in effect serving as energy reservoirs. In doing so, they can be used to realize dissipative quantum preparation and transport schemes for a variety of states. As such, this makes them a natural choice as the environment for the dissipation induced quantum transport scheme described in this paper. We leave the issue of the physical realization of this quantum transport scheme, as well as other possible issues arising from the formulation of the scheme, for future work. 
\section*{Acknowledgements}

This work is supported by a grant from the National Research Council of the Philippines as NRCP Project no. P-022. B.M.B. and V.P.V. acknowledge support from the Department of Science and Technology (DOST) as DOST-ASTHRDP scholars. \\

\appendix

\section{Derivation of the Master Equation}

To obtain Eq. \ref{masteqdyn}, we first evaluate the double commutator $\left[\mathrm{H}_{SB}(t),\left[\mathrm{H}_{SB}(s),\sum_{j}\rho_{S}(t)\otimes\ket{j}\bra{j}\otimes\rho_{B}\right]\right]$. Substituting equations \ref{timeevohamtot} and \ref{densmattime} into this expression, we obtain the following:

\begin{align}
&\left[\mathrm{H}_{SB}(t),\left[\mathrm{H}_{SB}(s),\sum_{j}\rho_{S}(t)\otimes\ket{j}\bra{j}\otimes\rho_{B}\right]\right]=\nonumber\\
&\sum_{j=1}^{N}\sum_{n,n'}\sum_{k,k'}\mathrm{e}^{\frac{i}{\hbar}(\varepsilon_{n'}-E_{k'})t}\mathrm{e}^{-\frac{i}{\hbar}(\varepsilon_{n}-E_{k})s}g^{*}_{n'k'}g_{kn}\nonumber\\
&\times\left(\hat{a}^{\dagger}_{n'}\hat{a}_{n}\rho_{n,j}(t)\otimes\ket{j}\bra{j}\otimes\hat{b}_{k'}\hat{b}^{\dagger}_{k}\rho_{B}\right.\nonumber\\
&\left.-\hat{a}_{n}\rho_{n,j}(t)\hat{a}^{\dagger}_{n'}\otimes\ket{j+1}\bra{j+1}\otimes\hat{b}^{\dagger}_{k}\rho_{B}\hat{b}_{k'}\right.\nonumber\\
&\left.-\hat{a}^{\dagger}_{n'}\rho_{n,j}(t)\hat{a}_{n}\otimes\ket{j}\bra{j}\otimes\hat{b}_{k'}\rho_{B}\hat{b}^{\dagger}_{k}\right.\nonumber\\
&\left.+\rho_{n,j}(t)\hat{a}_{n}\hat{a}^{\dagger}_{n'}\otimes\ket{j+1}\bra{j+1}\otimes\rho_{B}\hat{b}^{\dagger}_{k}\hat{b}_{k'}\right)\nonumber\\
&+\sum_{j=1}^{N}\sum_{n,n'}\sum_{k,k'}\mathrm{e}^{-\frac{i}{\hbar}(\varepsilon_{n'}-E_{k'})t}\mathrm{e}^{\frac{i}{\hbar}(\varepsilon_{n}-E_{k})s}g_{n'k'}^{*}g_{kn}\nonumber\\
&\times\left(\hat{a}_{n'}\hat{a}^{\dagger}_{n}\rho_{n,j}(t)\otimes\ket{j+1}\bra{j+1}\otimes\hat{b}^{\dagger}_{k'}\hat{b}^{\dagger}_{k}\rho_{B}\right.\nonumber\\
&\left.-\hat{a}^{\dagger}_{n}\rho_{n,j}(t)\hat{a}_{n'}\otimes\ket{j}\bra{j}\otimes\hat{b}_{k}\rho_{B}\hat{b}^{\dagger}_{k'}\right.\nonumber\\
&\left.-\hat{a}_{n'}\rho_{n,j}(t)\hat{a}^{\dagger}_{n}\otimes\ket{j+1}\bra{j+1}\otimes\hat{b}^{\dagger}_{k'}\rho_{B}\hat{b}_{k}\right.\nonumber\\
&\left.+\rho_{n,j}(t)\hat{a}^{\dagger}_{n}\hat{a}_{n'}\otimes\ket{j}\bra{j}\otimes\rho_{B}\hat{b}_{k}\hat{b}^{\dagger}_{k'}\right).\nonumber\\
\label{doublecomm}
\end{align}
We then take the trace of Eq. \ref{doublecomm} over the environment variables, noting that $\mathrm{Tr}\left(\hat{b}^{\dagger}_{k'}\hat{b}_{k}\rho_{B}\right)=0$ and $\mathrm{Tr}\left(\hat{b}_{k'}\hat{b}^{\dagger}_{k}\rho_{B}\right)=\delta_{k',k}$. Thus, we obtain the following expression:

\begin{align}
&\mathrm{Tr}_{B}\left[\mathrm{H}_{SB}(t),\left[\mathrm{H}_{SB}(s),\sum_{n,j}\rho_{n,j}(t)\otimes\ket{j}\bra{j}\otimes\rho_{B}\right]\right]=\nonumber\\
&\sum_{j=1}^{N}\sum_{n,n'}\sum_{k}\mathrm{e}^{\frac{i}{\hbar}(\varepsilon_{n'}-E_{k})t}\mathrm{e}^{-\frac{i}{\hbar}(\varepsilon_{n}-E_{k})s}g^{*}_{n'k}g_{kn}\nonumber\\
&\times\left(-\hat{a}_{n'}\rho_{n,j}(t)\hat{a}^{\dagger}_{n}\otimes\ket{j}\bra{j}+\rho_{n,j}(t)\hat{a}^{\dagger}_{n}\hat{a}_{n'}\otimes\ket{j+1}\bra{j+1}\right)\nonumber\\
&+\sum_{j=1}^{N}\sum_{n,n'}\sum_{k}\mathrm{e}^{-\frac{i}{\hbar}(\varepsilon_{n'}-E_{k})t}\mathrm{e}^{\frac{i}{\hbar}(\varepsilon_{n}-E_{k})s}g_{n'k}^{*}g_{kn}\nonumber\\
&\times\left(\hat{a}^{\dagger}_{n'}\hat{a}_{n}\rho_{n,j}(t)\otimes\ket{j+1}\bra{j+1}-\hat{a}_{n}\rho_{n,j}(t)\hat{a}^{\dagger}_{n'}\otimes\ket{j}\bra{j}\right).\nonumber\\
\label{doublecommtrenvo}
\end{align}
We then make use of the boundary condition $\ket{N+1}=\ket{1}$, and in doing so we can factor out the position matrices $\ket{j}\bra{j}$ in the double commutator. Next, we integrate the double commutator over the time variable $s$, and simplify the resulting expression. This results in the following expression:

\begin{align}
&\int_{0}^{t}ds\;\mathrm{Tr}_{B}\left[\mathrm{H}_{SB}(t),\left[\mathrm{H}_{SB}(s),\sum_{j}\rho_{S}(t)\otimes\ket{j}\bra{j}\otimes\rho_{B}\right]\right]=\nonumber\\
&\sum_{j=1}^{N}\sum_{n,n'}\sum_{k}\frac{i\hbar}{\varepsilon_{n}-E_{k}}\left(\mathrm{e}^{-\frac{i}{\hbar}(\varepsilon_{n}-E_{k})t}-1\right)\mathrm{e}^{\frac{i}{\hbar}(\varepsilon_{n'}-E_{k})t}g^{*}_{n'k}g_{kn}\nonumber\\
&\times\left(-\hat{a}_{n'}\rho_{n,j+1}(t)\hat{a}^{\dagger}_{n}+\rho_{n,j}(t)\hat{a}^{\dagger}_{n}\hat{a}_{n'}\right)\otimes\ket{j}\bra{j}\nonumber\\
&-\sum_{j=1}^{N}\sum_{n,n'}\sum_{k}\frac{i\hbar}{\varepsilon_{n}-E_{k}}\left(\mathrm{e}^{\frac{i}{\hbar}(\varepsilon_{n}-E_{k})t}-1\right)\mathrm{e}^{-\frac{i}{\hbar}(\varepsilon_{n'}-E_{k})t}g_{n'k}^{*}g_{kn}\nonumber\\
&\times\left(\hat{a}^{\dagger}_{n'}\hat{a}_{n}\rho_{n,j}(t)-\hat{a}_{n}\rho_{n,j+1}(t)\hat{a}^{\dagger}_{n'}\right)\otimes\ket{j}\bra{j}.\nonumber\\
\label{doublecomminteg}
\end{align}
Next, we apply a rotating wave approximation to Eq. \ref{doublecomminteg}, wherein we set $\varepsilon_{n'}\approx\varepsilon_{n}$, and in doing so diagonalize the expression in $n$. In doing so, we can simplify Eq. \ref{doublecomminteg}, giving us the following equation:

\begin{align}
&\int_{0}^{t}ds\;\mathrm{Tr}_{B}\left[\mathrm{H}_{SB}(t),\left[\mathrm{H}_{SB}(s),\sum_{j}\rho_{S}(t)\otimes\ket{j}\bra{j}\otimes\rho_{B}\right]\right]=\nonumber\\
&\sum_{j=1}^{N}\sum_{n}\sum_{k}\frac{i\hbar}{\varepsilon_{n}-E_{k}}\left(1-\mathrm{e}^{\frac{i}{\hbar}(\varepsilon_{n}-E_{k})t}\right)\left|g_{kn}\right|^{2}\nonumber\\
&\times\left(-\hat{a}_{n}\rho_{n,j+1}(t)\hat{a}^{\dagger}_{n}+\rho_{n,j}(t)\hat{a}^{\dagger}_{n}\hat{a}_{n}\right)\otimes\ket{j}\bra{j}\nonumber\\
&-\sum_{j=1}^{N}\sum_{n,n'}\sum_{k}\frac{i\hbar}{\varepsilon_{n}-E_{k}}\left(1-\mathrm{e}^{-\frac{i}{\hbar}(\varepsilon_{n}-E_{k})t}\right)\left|g_{kn}\right|^{2}\nonumber\\
&\times\left(\hat{a}_{n}\hat{a}_{n}\rho_{n,j}(t)-\hat{a}^{\dagger}_{n}\rho_{n,j+1}(t)\hat{a}^{\dagger}_{n}\right)\otimes\ket{j}\bra{j}.\nonumber\\
\label{masteqrhsalmost}
\end{align}
Finally, adding up both sums in Eq. \ref{masteqrhsalmost}, we obtain the master equation given by Eq. \ref{masteqdyn} with the coefficients defined as in Eq. \ref{masteqcoeffs}.

\end{document}